\documentclass[conference]{IEEEtran}
\IEEEoverridecommandlockouts
\usepackage{cite}
\usepackage{amsmath,amssymb,amsfonts}
\usepackage{algorithmic}
\usepackage{graphicx}
\usepackage{float}
\usepackage{textcomp}
\usepackage{xcolor}
\usepackage{xspace}
\usepackage{algorithm}
\usepackage{algorithmic}
\usepackage{amsmath}

\newcommand{\etal}{\textit{et al.}\xspace}

\newcommand{\eg}{\textit{e.g.,}\xspace}
\newcommand{\rev}[1]{\textcolor{black}{#1}}

\def\BibTeX{{\rm B\kern-.05em{\sc i\kern-.025em b}\kern-.08em
    T\kern-.1667em\lower.7ex\hbox{E}\kern-.125emX}}
\begin{document}

\title{A Vehicle Transmission Scheduling Scheme for Supporting Vehicle Trust Management}

\author{\IEEEauthorblockN{Qinwen Hu}
\IEEEauthorblockA{\textit{State Key Laboratory of Internet of Things for Smart City} \\
\textit{University of Macau}\\ China \\
qinwenhu@um.edu.mo}
\and
\IEEEauthorblockN{Wanqing Tu}
\IEEEauthorblockA{\textit{School of Computer Science} \\
\textit{The University of Auckland}\\
New Zealand \\
w.tu@auckland.ac.nz}
}

\maketitle

\begin{abstract}
Recent developments in advanced sensors, wireless communications and intelligent vehicle control technologies have enabled vehicles to detect traffic anomalies on the road and then notify surrounding vehicles to improve traffic safety. However, due to the high-speed movement of vehicles and the frequent topological changes between vehicles, it is difficult for vehicles to evaluate the credibility of received messages. Quite a lot of research effort has been carried out to establish various trustworthiness platforms. These studies mostly focus on how to enhance the accuracy of credibility evaluation, overlooking that the transmission performance may affect the quality of vehicle messages. In this paper, we aim to support the improvement of credibility evaluation in vehicle networks by enhancing the transmission experience of vehicles. The proposed solution utilizes the vehicle's trajectory information, detection range and a road-side unit (RSU) coverage to form a controlled number of detection zones, which guarantees that events can be detected and reported while limiting the number of transmission vehicles. Furthermore, our scheme takes account of vehicle credibility and interference ranges when selecting reporting vehicles, supporting the timely and reliable delivery of vehicles’ event reports when accidents occur. Our ns2 evaluation shows that our scheme can greatly reduce delay and loss rates of vehicle messages to help existing studies on accurate vehicle credibility evaluation.
\end{abstract}

\section{Introduction}
In vehicle networks, vehicles are equipped with various onboard sensors to collect and share road-related information (e.g., road conditions, traffic congestion, accidents) with an RSU. Such information helps vehicles and RSUs to be aware of any emergency as well as predict potential traffic congestion, improving transportation safety and efficiency. However, due to the high mobility of vehicles, an RSU receives contacts or messages from different vehicles overtime. Therefore, it is important for an RSU to accurately understand the credibility of different vehicles, supporting efficient traffic and road management. 

Much research effort has been spent on various algorithms, experiments, and technical standardisations in order to enhance the trustworthiness of vehicle reports. In general, they can be classified as centralised solutions or distributed solutions. Centralised solutions （\eg \cite{shen2019cate,wang2019neighborhood}), employ central servers to collect feedback or reports from vehicles. With vehicles' reports, central servers can then make action decisions based on vehicles' credibility. For distributed solutions (\eg \cite{chen2013beacon,yang2018blockchain,cheng2019trust}), they allow vehicles to send feedback or reports to each other. The trustworthiness of these reports is evaluated by vehicles based on the credibility and the locations of reporting vehicles. Vehicles may also update the credibility of other vehicles based on the evaluation results of those received reports. These related studies mainly focus on efficiently evaluating and dynamically updating vehicle credibility ratings by referring to received reports/feedback and historic vehicle behaviour. Most of them assume that vehicle reports are received accurately and in real time for making decisions and updating credibility. 

In a vehicle network, multiple adjacent vehicles may detect events or accidents on the road simultaneously. When they report such detection, the transmissions of their reports may interfere with each other due to the wireless broadcast property. Such interference easily causes packet loss, data errors, or long transmission delays \cite{ ntaganda2018bit,schmidt2013bit,li2019efficient,chang2018distributed }, negatively affecting the report quality received at receivers. In another words, the trustiness of vehicle reports not only relies on vehicles' credibility and behaviour but also is affected by transmission quality. In this paper, we investigate how to support accurate trust management in vehicle networks by improving vehicle transmission experience. In detail, we present the following contributions. 

\begin{itemize}
\item Formation of detection zones. The coverage of the RSU is divided into a controlled number of detection zones based on vehicles’ detection range. Detection zones are formed to limit the number of vehicles sending reports to an RSU, while ensuring these selected vehicles to be able to report events anywhere on the road.

\item Allocation and ranking of vehicles. Vehicles on the road are allocated to detection zones periodically based on their current locations, due to the frequent dynamicity of vehicles’ locations. Within each detection zone, vehicles are ranked in terms of their credibility. The ranking is also updated dynamically to reflect changes in vehicles’ detection zones.

\item Selection of reporting vehicles. Within each detection zone, a subset of vehicles is selected to report on any road events that may be detected. This selection refers to the ranking of vehicles inside a detection zone, as well as the distances between vehicles so as to control report transmission interference. These two factors are combined as a vehicle weight by which reporting vehicles are selected periodically to report road events to the RSU. 
\end{itemize}

Our NS2 simulation results show that our method greatly helps the selected schemes in terms of controlling report loss rates and average report delays when vehicles send feedback to the RSU. More specifically, in our simulations, our approach helps existing vehicle trust management schemes to achieve a report loss rate $<5\%$ with acceptable average report delays.

The rest of the paper is organized as follows. Section \ref{RW} gives a comprehensive overview of current literature related to trustworthiness schemes in vehicle networks. Section \ref{MT} presents our proposed mechanism. Section \ref{ER} describes the simulation methodology and performance evaluation between two selected solutions with and without our proposed solution. Finally, we make some concluding remakes in section \ref{Con}.

\section{Related work}
\label{RW}
As a fundamental topic, a trust management scheme is considered in every emerging technology. In the past five years, many researchers \cite{wang2019neighborhood,shrestha2018centralized,chen2013beacon,shafer1992dempster,javed2018trust,yang2018blockchain} paid attention to design a trustworthiness solution in vehicle networks. These solutions utilize vehicles, RSUs and cloud servers to monitor malicious behaviours. Wang \etal \cite{wang2019neighborhood} proposed a centralized trustworthiness solution. The cloud servers is utilized to collect, evaluate, and store the trust values of all vehicles. The central server is usually assumed to be a fully trusted entity. However, the centralized solution is not practical to cope with large numbers of vehicles, resulting in high latency. Besides, the single point of failure is another big challenge of using centralized solution. Shrestha \etal \cite{shrestha2018centralized} proposed a peer to peer verification scheme to reduce the computation and transmission delay caused by the centralized solution. Liao \etal \cite{liao2013trust} used an RSU to evaluate the credibility of vehicles based on the accuracy of their previous incident reporting (by comparing it with actual ground truth). The solution reduces the number of false positives and false negatives, but the solution has not solved the single point failed problem. Similarly, Huang \etal \cite{huang2017distributed} utilized the RSUs to construct a decentralized solution to remove the single point of failure problem, the distributed RSUs store the vehicels' credibility, and the vehicle can request the RSU to query the credibility of neighbouring vehicles. However, this design encounters a consistent data issue. Yang \etal \cite{yang2018blockchain} proposed a decentralized trust management system based on blockchain technology to address the consistent data issue. Yang’s solution aggregates all the received reports using specific algorithms. If the aggregated value exceeds a threshold, the receiver will trust the content of the data and update the credibility of all the reporting vehicle in the blockchain. 

The conventional solutions do not take  into account the impact quality of service (QoS) particularly in a dense traffic scenario \cite{tu2014multi,toghi2019spatio,peron2019efficient,schmidt2013bit,zhou2019lppa,tu2016efficient}, which may lead to a large transmission delay and a high report loss rate. This paper performs an experimental study of the proposed algorithm, we aim to improve the transmission quality in existing solutions by considering vehicle credibility and interference ranges when selecting reporting vehicles.
\section{The Vehicle Transmission Scheduling Algorithm (VTS)}
\label{MT}
\rev{Denote a RSU as $R$}. In our system, instead of asking all vehicles to report road events, $R$ selects a subset of vehicles in its coverage to report road events. The selection of this subset of vehicles is not trivial as it is hard to predict where on a road an event would happen and the accurate detection distance of a vehicle is limited. Our idea is to divide $R$'s coverage into \rev{a controlled number of detection zones that vehicles are dynamically allocated to. Then, within each detection zone, the vehicle selection scheme is designed to select event reporting vehicles based on their rankings.} This ensures that events can be detected and reported no matter where on the road that events take place.

\subsection{Formation of Detection Zones}
\label{detectionzone}
\rev{The formation of detection zones (DZs) is implemented by $R$. In order to allow vehicles inside a DZ to accurately sense events in this zone, $R$ forms DZs by referring to the accurate detection distance of the vehicles. Meanwhile, in order to form a controlled number of DZs within the coverage of $R$, our detection zones are in the shape of cubes. This is because cubes not only may fully cover $R$'s coverage without overlaps but also} require relatively simple calculations as compared to other shapes.

\begin{figure}[ht]
\begin{center}
\includegraphics[scale=0.45]{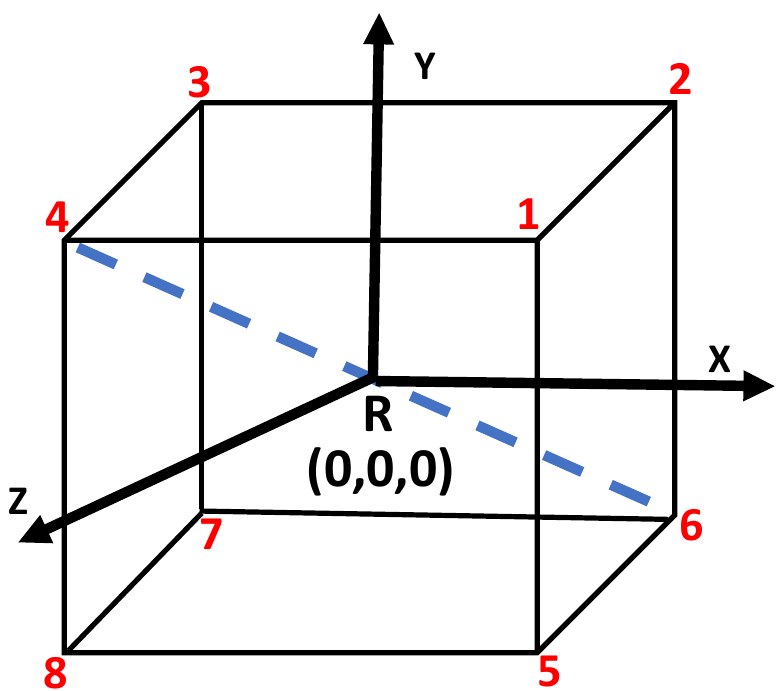}
\caption{An example of forming detection zones.}
\vspace{-8mm}
\label{zones}
\end{center}
\end{figure}

\rev{Without loss of generality, assume $R$ locates at (0,0,0) and the accurate detection distance of vehicles in our system is $d$. $R$ forms the first DZ by regarding itself as the central point of the cube. Then, in order to guarantee any vehicles inside this DZ can accurately sense events in this zone, the longest distance in this DZ (i.e., the length of a diagonal line (shown by the blue dotted line in Fig.\ref{zones})) should be $\leq{d}$. To form the DZ as large as possible, helping to reduce the number of DZs that need to form in $R$'s coverage, we use $d$ as the length of a diagonal line in this DZ. Therefore, the sides of the DZ should have the length $s$ of the sizes of a DZ is calculated as $$(\sqrt{2s})^2+s^2=d^2\Rightarrow{s=\frac{d}{\sqrt{3}}}.$$}

\rev{With the side value of the DZ, it is not hard to obtain the eight vertices as illustrated by red numbers in Fig.\ref{zones}. More specifically, the coordinates for vertex 1 are $(\frac{d}{2\sqrt{3}},\frac{d}{2\sqrt{3}},\frac{d}{2\sqrt{3}})$, for vertex 2 are $(\frac{d}{2\sqrt{3}},-\frac{d}{2\sqrt{3}},\frac{d}{2\sqrt{3}})$, for vertex 3 are $(-\frac{d}{2\sqrt{3}},-\frac{d}{2\sqrt{3}},\frac{d}{2\sqrt{3}})$, for vertex 4 are $(\frac{d}{2\sqrt{3}},-\frac{d}{2\sqrt{3}},\frac{d}{2\sqrt{3}})$,
for vertex 5 are $(\frac{d}{2\sqrt{3}},\frac{d}{2\sqrt{3}},-\frac{d}{2\sqrt{3}})$, for vertex 6 are $(\frac{d}{2\sqrt{3}},-\frac{d}{2\sqrt{3}},-\frac{d}{2\sqrt{3}})$, for vertex 7 are $(-\frac{d}{2\sqrt{3}},-\frac{d}{2\sqrt{3}},-\frac{d}{2\sqrt{3}})$ and for vertex 8 are $(-\frac{d}{2\sqrt{3}},\frac{d}{2\sqrt{3}},-\frac{d}{2\sqrt{3}})$ respectively. These eight vertices define the first DZ.} 

Once the first DZ is established, $R$ selects a face on this DZ to support the formation of further DZs. Without loss of generality, \rev{$R$ first selects the face having the vertices 1, 2, 3, and 4.} This next DZ is also a cube sharing the selected face with the first DZ. Therefore, the other four vertices of the second DZ, is  $(\frac{d}{2\sqrt{3}},\frac{d}{2\sqrt{3}},\frac{d}{2\sqrt{3}}+\frac{d}{\sqrt{3}})$, $(\frac{d}{2\sqrt{3}},-\frac{d}{2\sqrt{3}},\frac{d}{2\sqrt{3}}+\frac{d}{\sqrt{3}})$, $(-\frac{d}{2\sqrt{3}},-\frac{d}{2\sqrt{3}},\frac{d}{2\sqrt{3}}+\frac{d}{\sqrt{3}})$, $(\frac{d}{2\sqrt{3}},-\frac{d}{2\sqrt{3}},\frac{d}{2\sqrt{3}}+\frac{d}{\sqrt{3}}).$ \rev{The centre of the face formed by these four new vertices can be obtained as} $(0,0, \frac{d}{2\sqrt{3}}+\frac{d}{\sqrt{3}})$. We call the Euclidean distance between $R$ and this centre as the distance covered by the two DZs (denoted as $d'$). If $r$$\geq$$d'$, where $r$ is the radius of R’s coverage, $R$ forms a further detection zone sharing the face defined by the four vertices that are not on the first DZ. 

\rev{$R$ then starts forming a new DZ sharing the face defined by the vertices 1, 2, 5, and 6 on the first DZ, by the similar procedure as above. $R$ continues forming further DZs until the distance covered by all DZs formed in this second procedure is $\geq{r}$. Then, similarly, $R$ starts forming a new DZ sharing the face defined by the vertices 1, 4, 5, and 8 on the first DZ. If the distance covered by the two DZs is $<r$, $R$ continues forming further DZs until the distance covered by all DZs in this third procedure is $\geq{r}$. Using the similar way above, $R$ forms more DZs by referring to other five faces on the first DZ.} 

For all the space inside these \rev{established DZs}, We call the space inside established DZs as covered space. Once the DZs along the eight faces of the first DZ have been formed, $R$ establishes other DZs in order to extend the covered space to the whole coverage of $R$. Without loss of generality, $R$ uses the second DZ as the reference zone to continue establishing further DZs along those faces on this reference zone that do not have adjacent DZs. If all covered space by these DZs does not fulfils $R$'s coverage, $R$ uses the third established DZ as the reference zone to form more DZs. The process continues until the covered space of all DZs fills $R$’s coverage. Then, the formation of detection zones completes. 
\subsection{Allocation and Ranking of Vehicles in Each Detection Zone}
\label{selectedlist}
Once $R$ has divided its coverage into a number of DZs, the next step is to allocate vehicles on the road to a DZ based on the vehicle’s location corrdinates. Like most vehicle networks, in our system, vehicles periodically send Hello messages. A Hello message generated by our vehicles contains three fields: vehicle ID, the coordinates of vehicle's current location, and vehicle's speed. A vehicle can may obtain its current location coordinates by global positioning system (GPS) \cite{murakami1999can}. $R$ extracts the vehicle's current location coordinates from the periodical Hello messages.

\rev{Suppose a vehicle's current location is (x,y,z), $R$ allocates this vehicle to a specific DZ by comparing (x,y,z) with DZs' coordinate boundaries. More specifically, for the $i$th DZ ($i\in{[0,n-1}$), where $n$ is the total number of DZs in R's coverage, we let  $x_{i,min}$, $y_{i,min}$, and $z_{i,min}$ be the minimum values of all points in this DZ on the x, y and z axes, respectively, and $x_{i,max}$, $y_{i,max}$, and $z_{i,max}$ be the maximum values. $R$ compares $x$ with $x_{i,min}$ and $x_{i,max}$, $y$ with $y_{i,min}$ and $y_{i,max}$, and  $z$ with $z_{i,min}$ and $z_{i,max}$. The vehicle is allocated to the $i$th DZ only when the vehicle's coordinates meet $x_{i,min}\leq{x}\leq{x_{i,max}}$, $y_{i,min}\leq{y}\leq{y_{i,max}}$, and $z_{i,min}\leq{z}\leq{z_{i,max}}$. The relationship between a vehicle and a DZ is dynamic as the vehicle keeps moving. By the above way, based on the latest Hello messages, $R$ dynamically maps vehicles to DZs.}

For vehicles allocated to the same DZ, $R$ forms a ranking list by sorting all vehicles in descending order of their credibility. Many vehicle trust management schemes (\eg \cite{yang2018blockchain,wang2019neighborhood}) have been designed to maintain vehicles' credibility. In \cite{yang2018blockchain},the blockchain is used to maintain the vehicles' credibility across RSUs. In \cite{wang2019neighborhood}, the cloud server is utilized to collect, calculate, and store the trust values of all vehicles. Our scheme can be easily integrated with these existing studies to enable $R$ to obtain vehicles' credibility. As we will introduce in the next subsection, this ranking list is referred to when selecting vehicles to report events on the road. Our ranking list dynamically changes with vehicles enters or leaves a DZ as below.
\begin{enumerate}
  \item \rev{When $R$ receives a Hello message with a vehicle ID which is not currently on the ranking list for the DZ, $R$ regards the vehicle as a new one entering into the DZ. $R$ achieves the credibility of this vehicle from the blockchain or other vehicles (depending on the employed vehicle trust management scheme). $R$ then places the ID of this vehicle on the ranking list so that all vehicles in the DZ are ranked in the descending order of their credibility.}
  \item \rev{When $R$ has not received a Hello message, within a time period (say 3 Hello periods),  from a vehicle whose ID is on the ranking list for the DZ, $R$ regards that the vehicle has left the DZ. The ID of the vehicle will hence be removed from the ranking list.} 
\end{enumerate}
Through the above steps, $R$ complete the process of ranking the vehicles in each DZ based on their credibility.
\subsection{Selection of Reporting Vehicles}
\label{SelectedVehicles}
Once the $n$ vehicles belonging to the same DZ are ranked, $R$ selects $m$ $(m<n)$ vehicles as event reporting vehicles in this DZ. The first vehicle selected is the one on top of the ranking list. To select the remaining $(m-1)$ reporting vehicles, $R$ not only refers to vehicles' credibility but also takes into account whether vehicles interfere with those selected reporting vehicles. In another words, $R$ assigns each vehicle that have not been selected as a report vehicle a weight. The weight of the $i$th non-reporting vehicle is expressed by 
$$
\omega_i=\alpha\times{C_i}+\sum_{j=0}^{l-1}(\beta_{i,j}\times{I_{i,j}}),
$$
where $i\in{[0,n-l-1]}$, $l$ is the number of currently selected reporting vehicles, $C_i$ is the credibility of the $i$th non-reporting vehicle, $I_{i,j}$ indicates the interference relationship between the $i$th non-reporting vehicle and the $j$th reporting vehicles, if the $i$th non-reporting vehicle may interfere with the $j$th reporting vehicles, $l_{i,j}=0$; otherwise, $I_{i,j}=1$. Besides, $\alpha$ and $\beta_{i,j}$ are $\in{(0,1)}$ with $\alpha+\sum_{j=0}^{l-1}\beta_{i,j}=1$. 

The weights of a non-overlapping vehicle may be changed when a new reporting vehicle is selected, as the interference relationship between this non-overlapping vehicle and the new reporting vehicle needs to be included in the weight expression. At each time when the weights of all non-reporting vehicles are updated, $R$ selects a non-reporting vehicle with the largest weight as a new reporting vehicle until $m$ reporting vehicles are selected. Then, $R$ selects $m$ vehicles belonging to another DZ as the reporting vehicle of this DZ, by the similar way above. Once all DZs have $m$ reporting vehicles selected, $R$ stops the current reporting vehicle selection procedure. 
$R$ selects $m$ reporting vehicles for all DZs periodically because vehicles change their DZs during the movement. When deciding $R$'s period of updating the reporting vehicles in all DZs, vehicles' driving speeds, the size of DZs, and the period of the Hello message should be referred to. 
 
\section{Simulation Evaluation}
\label{ER}
In this section, by using NS2.35, we apply our algorithm to two trust management schemes recently proposed for vehicle networks.
\begin{itemize}
\item The blockchain-based trust management (BCTM) \cite{yang2018blockchain}. It allows all vehicles which detect the accident to broadcast the incident report to the surrounding neighbours. Based on the received reports, each receiver aggregates the received reports and generate a trust rate for a specific event. 
\item The neighborhood-based trust management (NBTM) \cite{wang2019neighborhood}. It only allow the vehicles near a place where the accident has just happened to report the incident. When the RSU receives the reports, it requests the cloud server to verify the credibility of the reporting vehicle, and return the reporting vehicle’s credibility. The RSU will drop the report from untrusted vehicles and send the correct incident information to vehicles and RSUs around it.  
\end{itemize}
More specifically, we simulate the two schemes with and without our VTS algorithm. Namely, we evaluate BCTM-with-VTS, BCTM-without-VTS, NBTM-with-VTS, and NBTM-without-VTS along the following performance metrics. 
\begin{itemize}
\item Average report delays (ARD). The report delay refers to the period of time from a vehicle sending a report to the RSU receiving this report. The average report delay is calculated by $\frac{\sum_{i=0}^{m-1}RD_i}{m}$, where $m$ is the number of reporting vehicles in a detection zone and $RD_i$ is the report delay of the $i$th reporting vehicle.
\item Report loss rate (RLR). When more than one vehicle sends reports at the same time to the RSU, report loss may occur. RLR is calculated by $RLR=\frac{R_t-R_r}{R_t}$, where $R_t$ is the number of vehicles sending a report and $R_r$ is the number of reports received at the RSU.
\end{itemize}
The major parameters employed in our simulation is listed in Table \ref{tab1}.
\begin{table}[htbp]
\caption{NS2 Simulation Parameters}
\begin{center}
\begin{tabular}{|c|c|}
\hline
\textbf{Parameters}&{\textbf{Values}} \\
\hline
{MAC standard}&{802\_11} \\
\hline
{Antenna}&{OmniAntenna} \\
\hline
{The number of vehicles}&{50 vehicles \cite{yang2018blockchain}} \\
\hline
{Hello packet and event report packet size}&{100 bytes \cite{javed2018trust,etsi2014102}} \\
\hline
{The interval of Hello messages}&{100 ms \cite{etsi2014102}} \\
\hline
{Wireless bandwidth}&{1 Mbps \cite{kaaniche2011qos}} \\
\hline
{The transmission range of the RSU}&{1000 meters \cite{tomar2010rsu}} \\
\hline
{The transmission range of vehicles}&{250 meters \cite{tomar2010rsu}} \\
\hline
{The accurate detection range of vehicles}&{100 meters \cite{luo2010pedestrian}} \\
\hline
\end{tabular}
\label{tab1}
\end{center}
\end{table}

\begin{figure}[h]
\begin{center}
\advance\leftskip-3cm
\advance\rightskip-3cm
\includegraphics[keepaspectratio=true,scale=0.4]{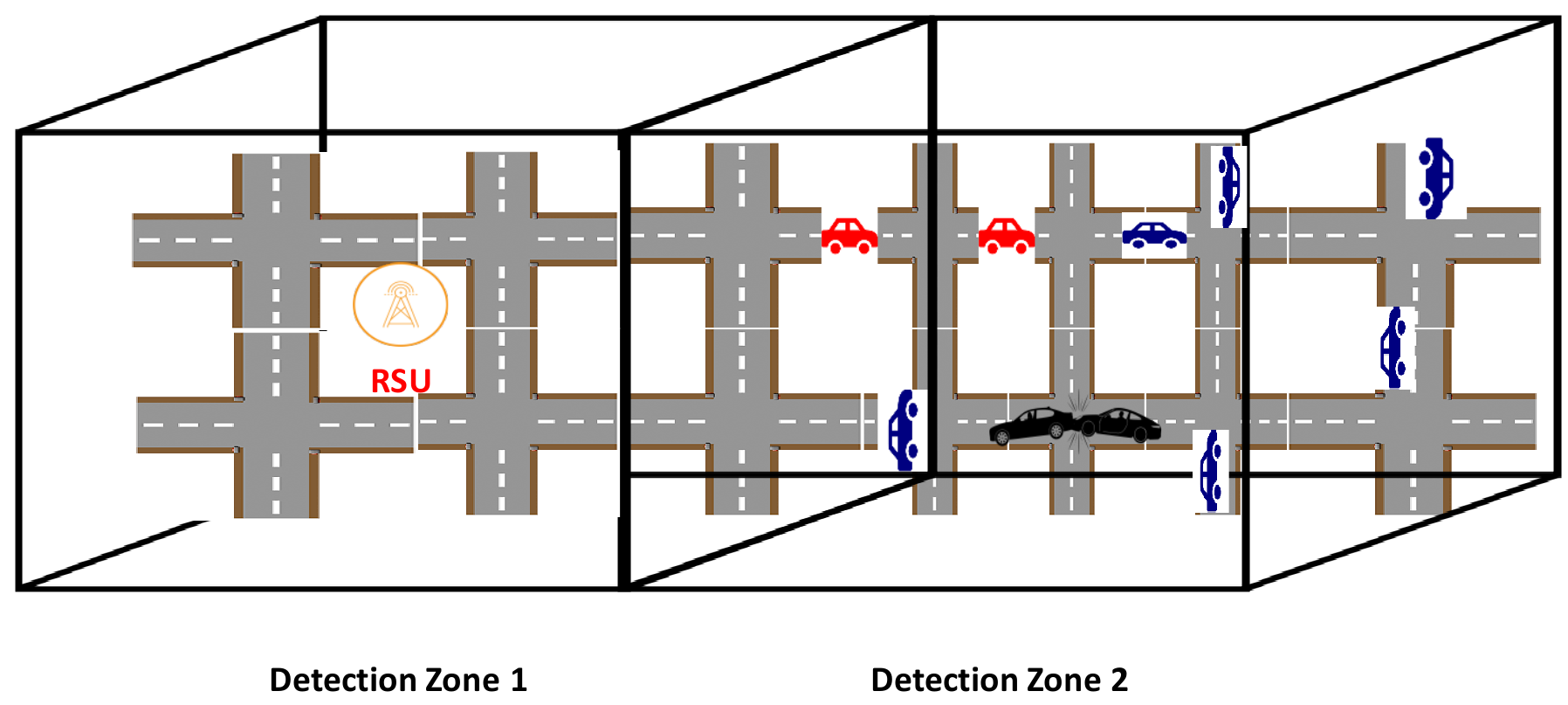}
\caption{The topology diagram of the proposed scheme.}
\vspace{-6mm}
\label{topo}
\end{center}
\end{figure}

Fig.\ref{topo} shows the topology diagram of this experiment. First, based on the detection zone formation approach in Section III A, we divided the RSU's coverage into 103 detection zones. For simplicity, we only show two detection zones in Fig.\ref{topo}. Then put the vehicle into the corresponding detection zone according to the current position of the vehicle. Suppose a car accident (two black vehicles in Fig.\ref{topo}) is found in the detection zone 2. 

\begin{figure}[h]
\begin{center}
\advance\leftskip-3cm
\advance\rightskip-3cm
\includegraphics[keepaspectratio=true,scale=0.45]{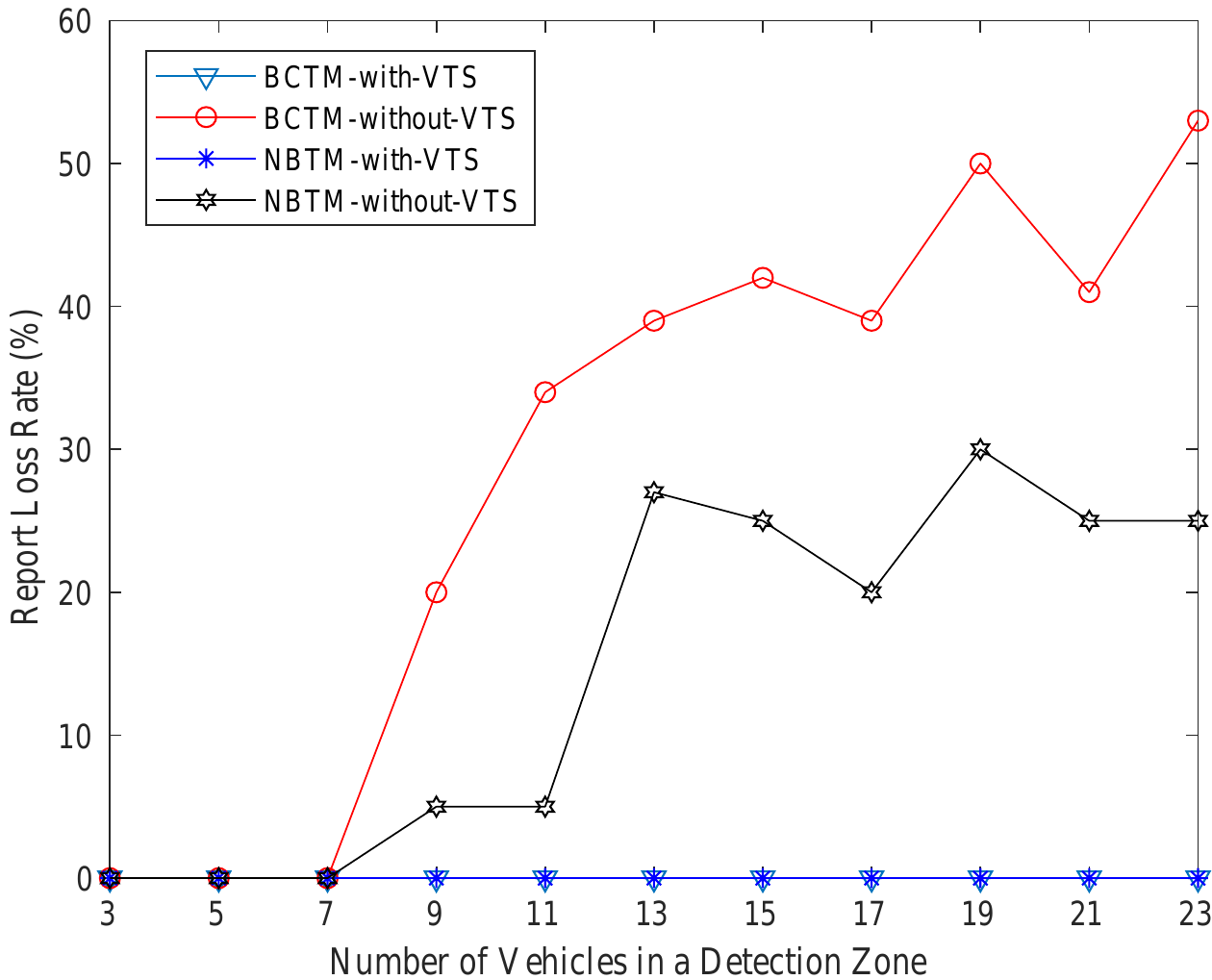}
\caption{RLR vs the number of selected reporting vehicles in a detection zone}
\vspace{-4mm}
\label{FN1}
\end{center}
\end{figure}
\paragraph{Report loss rate (RLR)} To study the impact of the number of reporting vehicles on RLR, for BCTM and NBTM, we configure the reporting vehicles to broadcast the report to all the surrounding vehicles and the RSU. While, for BCTM-with-VTS and NBTM-with-VTS, we select certain vehicles to unicast the report to the RSU. Fig.\ref{FN1} shows that as the number of reporting vehicles in detection zone 2 increases, RLR also increases if we use BCTM and NBTM. The reason is that the increase in the number of reporting vehicles will increase the number of interference domains, because vehicles are randomly located in different interference domains, resulting in the MAC layer collision occurs when broadcasting the report simultaneously. To reduce the RLR to 0\%, the RSU only select the reporting vehicles that are outside of the interference ranges of each other. Besides, not all vehicles in the detection zone are allowed to send reports, VTS selects those vehicles with high credibility, this design can reduce the occurrence of MAC layer collision. As a result, we observe that RLR in BCTM-with-VTS and NBTM-with-VTS remains 0\%.
\begin{figure}[h]
\begin{center}
\advance\leftskip-3cm
\advance\rightskip-3cm
\includegraphics[keepaspectratio=true,scale=0.45]{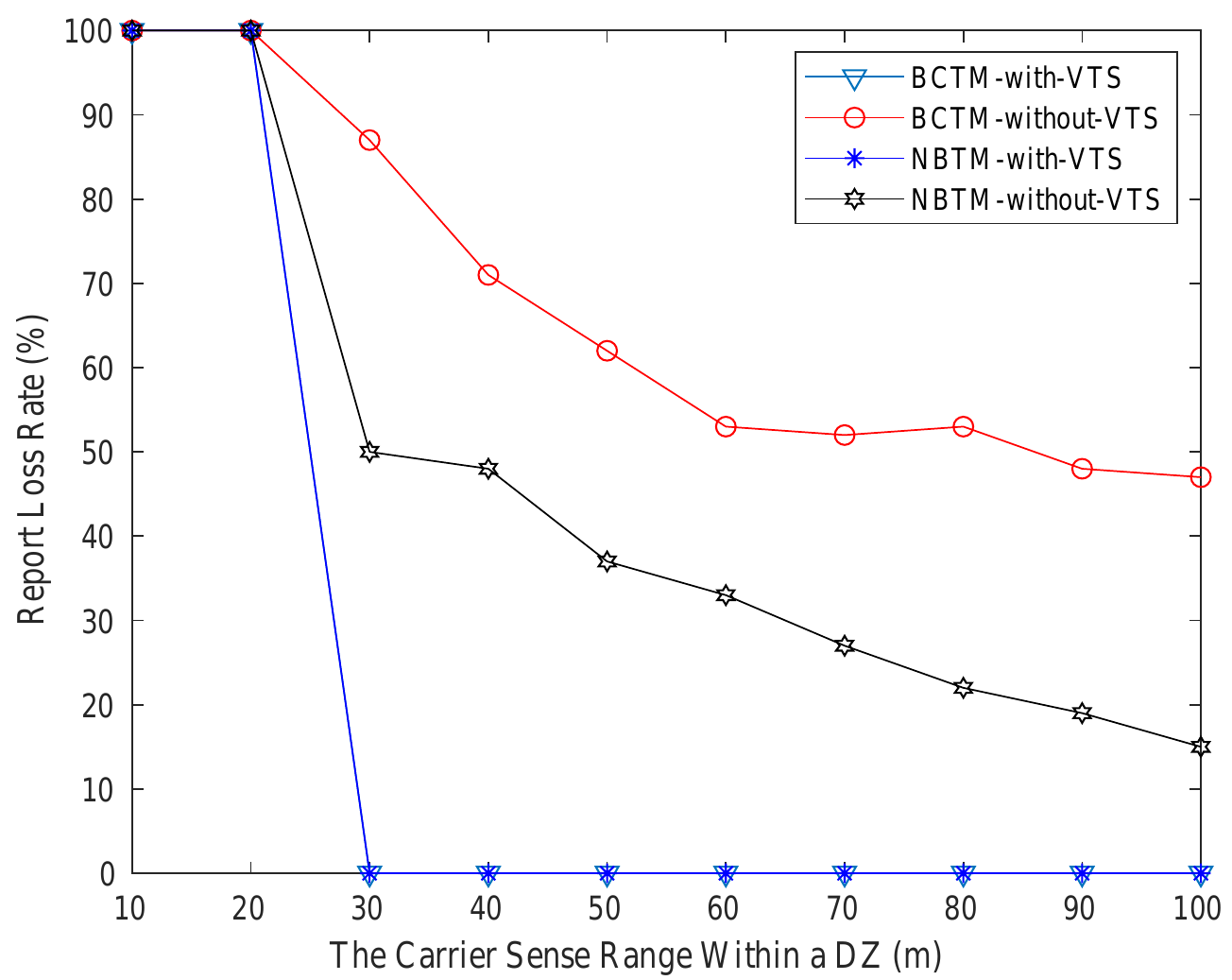}
\caption{RLR vs the carrier sense range in the detection zone.}
\vspace{-4mm}
\label{FNO}
\end{center}
\end{figure}

Fig.\ref{FNO} depicts how does the carrier sense range affect RLR, we assume that the detection zone 2 has 23 vehicles in total; the carrier sense range means that the nodes in the carrier sensing range can sense the sender’s transmission \cite{liu2007understanding}. Besides, it is used for the collision detection, before transmitting, a node first listens to the shared medium (such as listening for wireless signals in a wireless network) to determine whether another node is transmitting or not. If the carrier sensing range is configured very short, the hidden node problem may occur, a node may be transmitting where other nodes go undetected at this stage; as a result, the report is dropped due to the MAC layer collision. To check the impact of carrier sensing range on transmission and RLR, we have modified carrier sensing range in NS2 between 10 and 100 meters. The reason for choosing 100 meters is because we do not want to affect the transmission of other detection zones. Concerning the reality, the position of the vehicles is randomly distributed. For all the selected schemes, when carrier sense range is between 10 to 29 meters, the RSU cannot receive the report from the reporting vehicles due to the RSU is outside the reporting vehicle's carrier sense range, for this reason, the RSU cannot sense the sender’s transmission. After increasing the carrier sensing range, the RSU starts to get the reports. However, we found that because the reporting vehicles are randomly distributed, the vehicles may be located in different carrier sense areas, so the hidden terminal problem occurs. As a consequence, We found that RLR will decrease as the range of the carrier sense expands. For instance, the RLR in BCTM reduced from 90\% to 60\% and the RLR in NCTM dropped to 18\%. Due to VTS concerns the interference range between each reporting vehicles, so VTS can mitigate the hidden terminal and reduce RLR by selecting the reporting vehicles in the same sense area, Fig.\ref{FNO} depicts that as the carrier sense range increases, RLR in BCTM-with-VTS and NCTM-with-VTS remains at 0\% after 30 meters.

\paragraph{Average report delay (ARD)}
We try to investigate whether the number of vehicles in the detection zone 2 increases will affect ARD.
\begin{figure}[h]
\begin{center}
\advance\leftskip-3cm
\advance\rightskip-3cm
\includegraphics[keepaspectratio=true,scale=0.45]{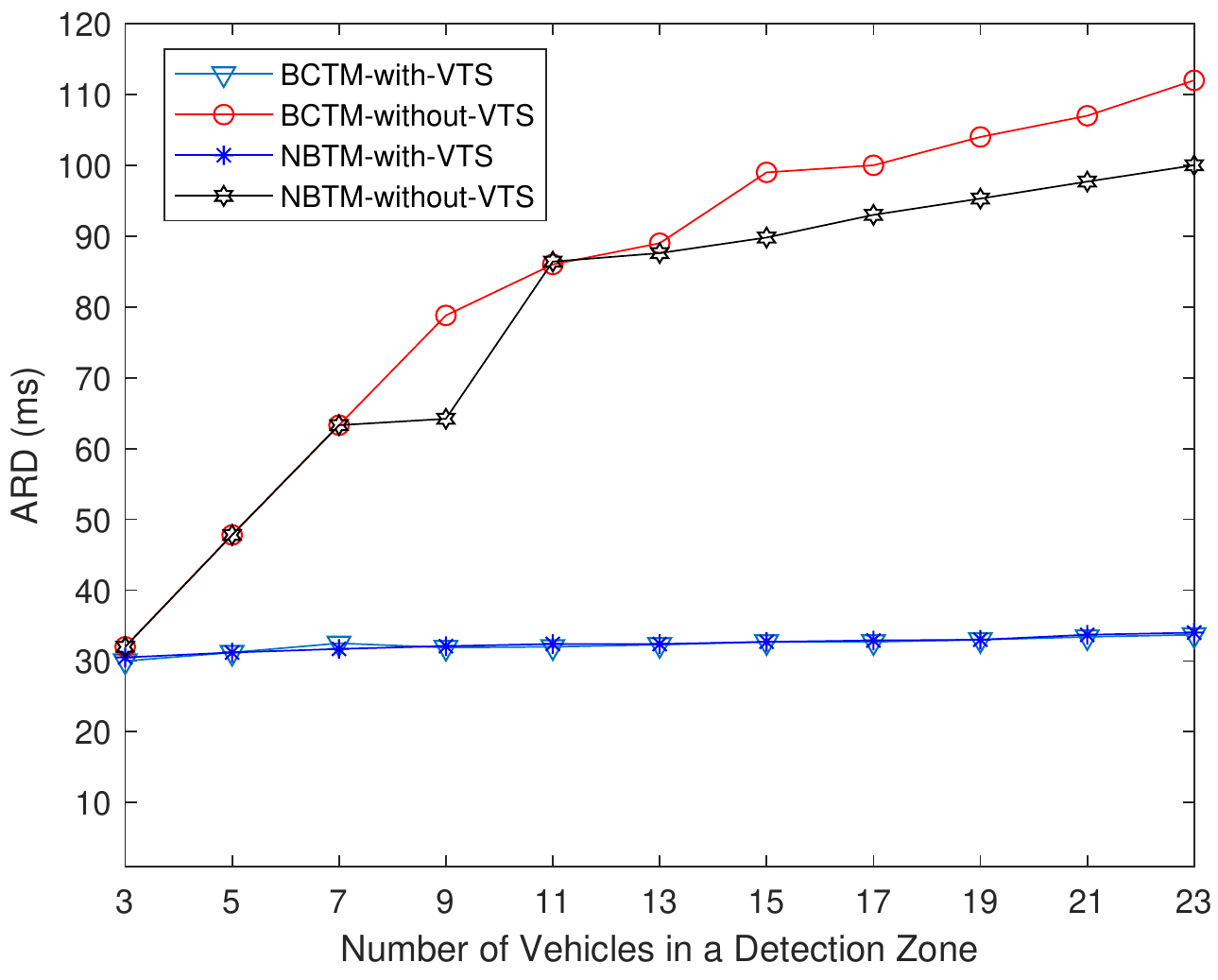}
\caption{ARD vs the number of selected reporting vehicles in a detection zone.}
\vspace{-4mm}
\label{ARD}
\end{center}
\end{figure}
Fig \ref{ARD} shows how ADR is affected by the number of reporting vehicles. For both BCTM and NBTM, as the number of vehicles increase, ARD gradually increased from 32ms to 112ms. The reason for this is that the reporting vehicles in the same carrier sense need to check the channel status if it senses the channel state is busy, then the reporting vehicle will wait until the line is idle for transmission, which will increase the transmission delay. Due to the content of the report is consistent, VTS considers reducing the number of reporting vehicles to accelerate the transmission speed. Thereby, the reporting vehicles do not need to wait long to obtain the channel resource. The simulation results show that ARD in BCTM-with-VTS and NBTM-with-VTS remains at 33ms when the reporting vehicles continue to increase.
\begin{figure}[h]
\begin{center}
\advance\leftskip-3cm
\advance\rightskip-3cm
\includegraphics[keepaspectratio=true,scale=0.45]{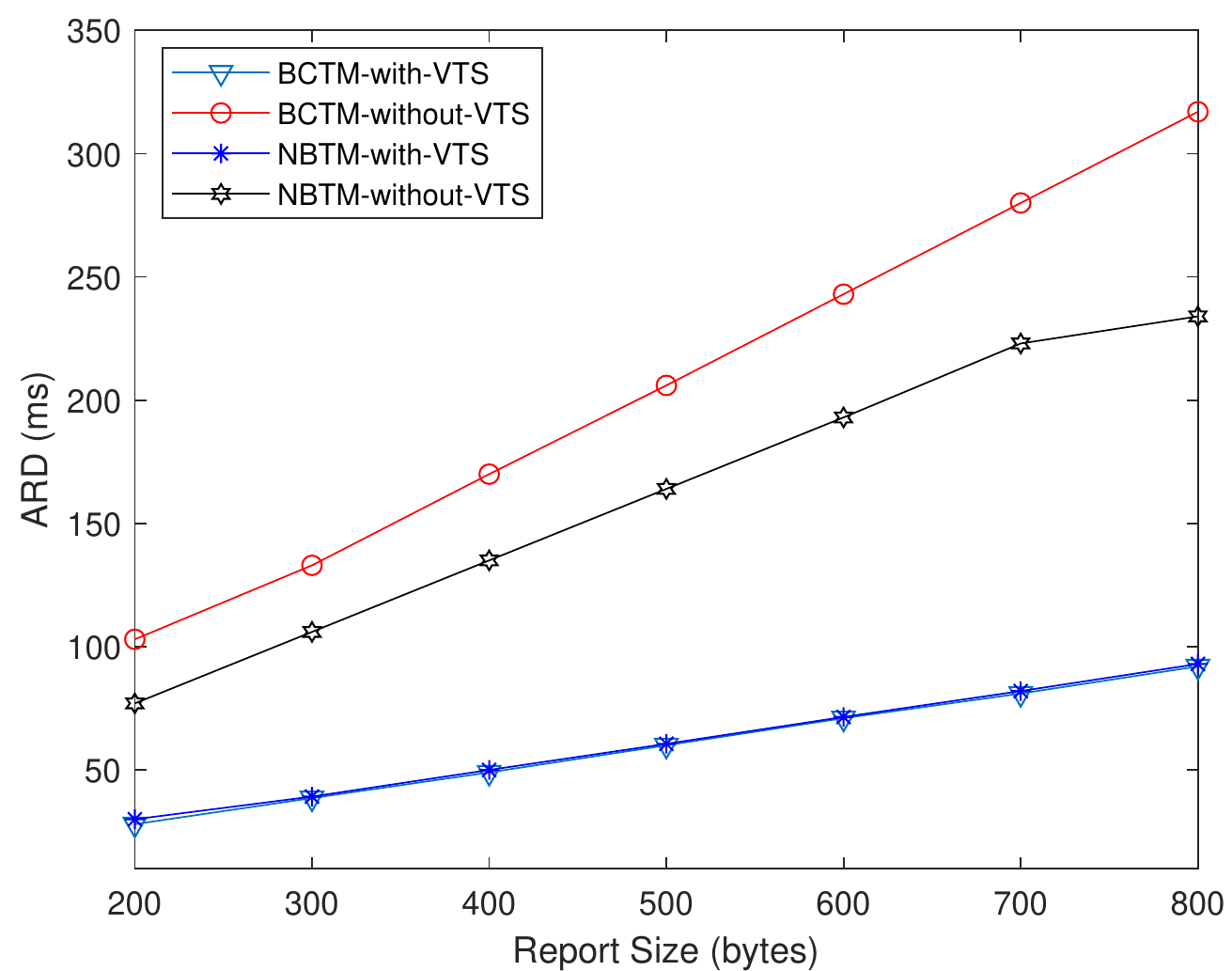}
\caption{ARD vs the safety report size.}
\vspace{-6mm}
\label{ARD1}
\end{center}
\end{figure}

In previous experiments, the vehicles send the report with a plain text. However, considering that security and privacy are getting more and more attention, the communication between RSUs and vehicles may adopt some security and privacy protection mechanisms. Raya and Hubaux \cite{raya2005security} concluded that based on the signing algorithms, the report could have a size between 200 and 800 bytes due to digital signatures and certificates. Hence, in this test, we investigate how does increasing the size of the report affect the ARD of the four selected methods scheme. Fig.\ref{ARD1} highlights that as a report size increases, the delay also increase drastically. For BCTM and NBTM schemes, due to many reporting vehicles send the same report simultaneously, the receiver takes a long time to aggregate all the reports together, such as, it takes 317ms to completely receive all reports with a report size of 800 bytes. Compared to the above two methods, considering that the content of the report is consistent, so VTS does not select all vehicles in the detection zone to send reports. Instead, VTS selects certain vehicles around the vehicles with the highest credibility in the detection zone. Although VTS still generate delays, but it has greatly reduced ARD from 317ms to 92ms when the report size is 800 bytes, this improvement ensures that other vehicles get the car accident information and respond to it immediately.

\section{CONCLUSION}
\label{Con}
A trustworthiness assessment is a key mechanism that can guarantee the security and reliability of information transmission in vehicle networks. Transmitting reports with minimum latency and low report loss rate in vehicle networks is the main issue that we undertake in this paper. We present a vehicle transmission scheduling algorithm for the vehicle trust management system. With the aid of this solution, the existing trustworthiness solutions can achieve low latency and low report loss rate when sending an incident report in the high-density area. The experiment results show that our vehicle transmission scheduling algorithm outperforms two selected solutions in terms of different metrics such as the ARD and RLR. As future work, we aim to study the optimal reporting vehicles selection mechanism and consider more parameters when selecting the reporting vehicle.

\bibliography{reference}
\bibliographystyle{IEEEtran}

\end{document}